\documentclass[aip,jcp,twocolumn,showpacs,superscriptaddress,10pt]{revtex4-1}
\usepackage{graphicx}
\usepackage{amsmath}
\usepackage{color}
\usepackage{multirow}
\begin{document}

\title{$\mathcal {C}-$IBI: Targeting cumulative coordination within an iterative protocol to derive coarse-grained models of (multi-component) complex fluids}

\author{Tiago E. de Oliveira}
\affiliation{Max-Planck Institut f\"ur Polymerforschung, Ackermannweg 10, 55128 Mainz Germany}
\affiliation{Universidade Federal do Rio Grande do Sul, Porto Alegre, Brazil}
\author{Paulo A. Netz}
\affiliation{Max-Planck Institut f\"ur Polymerforschung, Ackermannweg 10, 55128 Mainz Germany}
\affiliation{Universidade Federal do Rio Grande do Sul, Porto Alegre, Brazil}
\author{Kurt Kremer}
\affiliation{Max-Planck Institut f\"ur Polymerforschung, Ackermannweg 10, 55128 Mainz Germany}
\author{Christoph Junghans}
\email[]{junghans@lanl.gov}
\affiliation{Computer, Computational, and Statistical Sciences Division, Los Alamos National Laboratory, Los Alamos, NM 87545, USA}
\author{Debashish Mukherji}
\email[]{mukherji@mpip-mainz.mpg.de}
\affiliation{Max-Planck Institut f\"ur Polymerforschung, Ackermannweg 10, 55128 Mainz Germany}



\begin{abstract}
We present a coarse-graining strategy that we test for aqueous mixtures. 
The method uses pair-wise cumulative coordination as 
a target function within an iterative Boltzmann inversion (IBI) like protocol. 
We name this method coordination iterative Boltzmann inversion ($\mathcal {C}-$IBI). 
While the underlying coarse-grained model is still structure based and, thus, preserves pair-wise solution structure, 
our method also reproduces solvation thermodynamics of binary and/or ternary mixtures. 
Additionally, we observe much faster convergence within $\mathcal {C}-$IBI 
compared to IBI. To validate the robustness, we apply $\mathcal {C}-$IBI to study test cases of 
solvation thermodynamics of aqueous urea and a triglycine solvation in aqueous urea. 
\end{abstract}

\maketitle

\section{Introduction}

Systematic structural coarse-graining, or systematically reducing degrees of freedom of a complex (macro)molecular system, 
is a paramount challenge of multiscale modeling \cite{peter1,peter2,noid}. Deriving coarse-grained (CG) 
models has several advantages $-$ 1) When a multi-atom molecule and/or 
segments of a macromolecule are represented by a single site bead, the molecular dynamics (MD) simulation 
setups result in a smaller number of particles and thus give a significant computational gain. 2) The non-bonded interactions between CG beads 
are usually smooth. Therefore, large simulation time steps can be chosen. 3) The smooth interaction 
potentials lead to faster dynamics, which results in faster equilibration of the reference system.
In this context, there are several possible CG techniques of deriving CG potentials, such as force matching \cite{fm1,fm2}, inverse Monte Carlo \cite{imc,imc2}, 
Boltzmann inversion (BI) \cite{originibi1,originibi2} and its extension to (iterative) 
Boltzmann inversion (IBI) \cite{ibi}, relative entropy \cite{scott08jcp}, and/or potential of mean force \cite{harman09mac,shen11jctc,cp13jcp}. 
Additionally there are also well known CG models, examples include the Molinero water model \cite{molinero09jpcb} and the 
free energy based MARTINI model \cite{martini}. 
All these methods aim to target (or reproduce) a certain property of the underlying all-atom reference systems. 
Therefore, it is often difficult to map every property of a physical system within a unified CG model,
posing grand challenge in the representability and transferability of CG models \cite{peter1,peter2}. 
For example, in the case of liquid water, an IBI derived CG model usually presents a pressure of about 
6000 bars \cite{wang09epje}, which can be readjusted to 1 atm using a pressure correction \cite{wang09epje,ibi}.
However, this pressure correction compromises the fluid compressibility and thus results in unphysical fluctuations. In this context,
a more recent work, employing a pressure correction at barostat level, could preserve both pressure and compressibility within a 
unified CG model \cite{noid15jcp}. The complexity of deriving CG models grows even further when dealing with macromolecular 
solvation in solution mixtures, where thermodynamic properties are intimately linked to delicate intermolecular 
interactions and local concentration and/or conformational fluctuations \cite{mukherji13mac}. 

A widely used structure based CG method is the well known BI \cite{originibi1,originibi2} and the IBI \cite{ibi}, 
where the pair-wise non-bonded potential is obtained by inverting ${\rm g}(r)$ within 
an iterative procedure. In this context, being a simplified method, IBI works exceedingly well for several
systems, including polymer melt \cite{originibi1,originibi2,ibi}, single component fluids \cite{votca} and, also, to some extent for multicomponent fluids, 
to name a few. However, IBI does not guarantee that the derived CG model reproduces the same solvation thermodynamic state point 
as that of the reference all-atom system, especially for multi-component fluids. This is particularly because 
IBI targets to fit ${\rm g}(r)$ and, for binary mixtures, the convergence of pair-wise ${\rm g}(r)$ 
(unity at large distances) often suffers from the very nature of CG protocol. 
Therefore, a small absolute deviation in ${\rm g}(r)$ can lead to a significant error in the 
cumulated coordination numbers. For example, estimation of coordination, given by
\begin{equation}
{\mathcal C}_{ij}(r) = 4\pi \int_0^r {\rm g}_{ij}(r') r'^2 dr'
\label{eq:kbi}
\end{equation}
with the indices $i$ and $j$ standing for every set of pairs, 
uses a volume integral of ${\rm g}(r)$. This requires ${\rm g}(r)$ to 
be multiplied by a factor of $4\pi r^2$ and a small error in ${\rm g}(r)$ 
are weighted by a factor of $4\pi r^2$. Therefore, it is important to 
obtain a precise estimate of ${\rm g}(r)$ for all $r$ values and 
thus presents a need to devise a better, yet simple, CG protocol, which is the motivation behind this work.

Additionally, an accurate, yet simple, CG model is highly important for hybrid simulations, such as the adaptive 
resolution scheme (AdResS) \cite{kremer05jcp,kremer12prl}. In AdResS a small all-atom region is coupled to a CG reservoir.
Correct thermodynamic conditions within the all-atom region are strongly related to the particle fluctuations 
and thus requiring a CG region that presents precise measure of the fluctuations compared to the all-atom region.
This is even more important for the multicomponent fluids \cite{mukherji13mac,mukherji12jctclet}.  

The above mentioned reasoning poses grand challenges to the derivation of CG models to study solvation properties of solvent mixtures, 
especially because solvation thermodynamics is dictated by $-$ 1) the energy density within the solvation volume of the macromolecule, 
2) the local concentration fluctuation of the two solvent components, and 3) the entropic contributions, especially near the 
transition region of macromolecules where a delicate balance between entropy and energy plays a key role. 
In this context, the energy density is not only related to the (co)solvent-macromolecule 
interaction strengths, but also to the solution composition within the solvation volume \cite{mukherji13mac} and thus is related to 
the first shell coordination number. However, fluctuations are related to the convergence of the tails of 
pair-wise radial distribution functions ${\rm g}(r)$ \cite{kb51jcp}. 
This presents a need for a protocol that can get both the above 
scenarios correct within a simplified CG strategy. Therefore, in this work we devise a method that aims to 
use ${\mathcal C}(r)$, as a target function within an IBI-like iterative protocol. 
Our method not only gives a precise estimate of the coordination 
number in comparison to the reference all-atom system, but also the precise estimate of the solvation properties.
Added advantage of this protocol is that it presents a much faster convergence in comparison to the 
conventional IBI protocol.

The remainder of the paper is organized as follows: in section \ref{method} we sketch the method followed by the 
result and discussion in section \ref{res}. Finally in section \ref{sec:con} we draw our conclusions.

\section{Method and model}
\label{method}

\subsection{All-atom simulations}

The CG model is derived from an underlying all-atom reference system. We use test cases of aqueous urea mixtures and 
the solvation of a single triglycine in aqueous urea mixtures, which was studied by two of us in an earlier 
work \cite{mukherji12jctc}. The reference all-atom simulations are performed using GROMACS \cite{gro}. 
We use the Kirkwood-Buff derived force field for urea \cite{smith03jpcb} 
and the SPC/E water model \cite{spce}. A combination of these two force-fields for the aqueous urea mixtures are 
known to reproduce correct solution thermodynamics. We consider four different urea molar concentrations $c_{u}$, ranging 
between 2.0 $-$ 8.0 M. We restrict the concentration to below 8.0 M because urea is known to
denature proteins at around 6 M solutions \cite{afinsen}. System sizes are chosen to be consisting of $\sim 16000$ molecules, 
where we consider four different mole fractions 2.0 M, 4.0 M, 6.0 M and 8.0 M.
The specific choice of these system sizes give reasonable convergence in the thermodynamic properties, which usually 
suffer from severe system size effects within small systems \cite{mukherji13mac,mukherji12jctclet}.
The force field parameters for triglycine are taken from Gromos43a1 \cite{groms}. The all-atom simulations are performed 
for 25 ns within an NpT ensemble, where the pressure is controlled with a Berendsen barostat at 
1 atm pressure with a coupling time of 0.5 ps \cite{berend}. The initial configurations for the all-atom 
simulations are taken from a 50 ns long equilibrated sample from our earlier study \cite{mukherji12jctc}. 
The temperature is set to 300 K using a Berendsen thermostat with a coupling time of 0.1 ps. The integration time step is 1 fs. The interaction cutoff is chosen 
as 1.4 nm. Electrostatics is treated using particle mesh ewald \cite{pme}. The bond lengths of the urea molecules and 
triglycine is constrained using the LINCS algorithm \cite{lincs}.

\subsection{Coarse-grained simulations}

The IBI and $\mathcal C-$IBI derived CG potentials are used to simulate full blown CG configurations. 
The temperature is set to 300 K using a Langevin thermostat with a damping constant of 0.2 ps.
Simulation time step is chosen as 4 fs and the cutoff distance is 1.4 nm.
Simulations are conducted for 50 ns. We use the last 25 ns of a trajectory from 50 ns to calculate 
observables, such as ${\rm g}(r)$, urea activity coefficients $\gamma_{uu}$ and 
the shift in solvation free energy $\Delta \mathcal{G}_{t}$ of triglycine. CG simulations are 
also performed using GROMACS \cite{gro}.

\section{Results and discussions}
\label{res}

\subsection{$\mathcal C-$IBI: Coordination iterative Boltzmann inversion}

\begin{figure}[ptb]
\includegraphics[width=0.46\textwidth,angle=0]{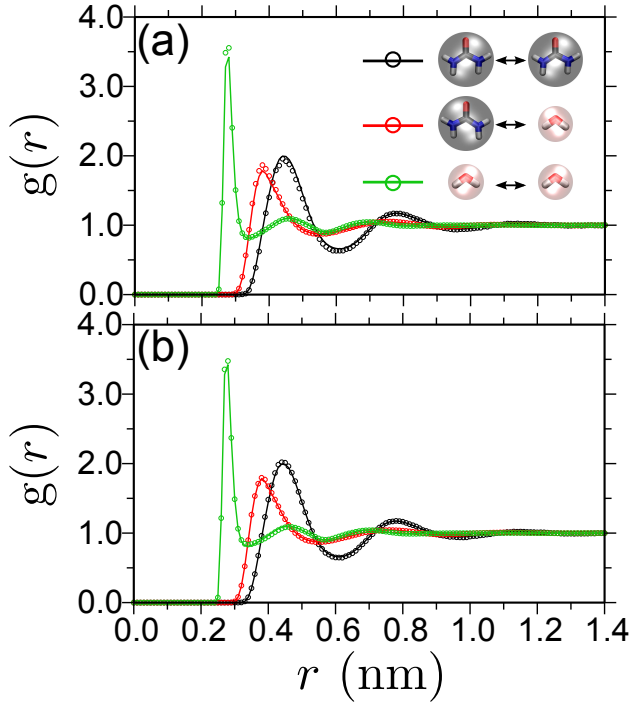}
\caption{Pair distribution function ${\rm g(r)}$ between different solvent pairs for a 6 M aqueous urea mixture. The solid lines 
present the reference all-atom data and the symbols are fitted coarse-grained ${\rm g(r)}$ after 25 iterations. Part (a) is the data 
corresponding to the iterative Boltzmann inversion (IBI) and part (b) corresponds to the coordination iterative Boltzmann inversion ($\mathcal C-$IBI). 
\label{fig:rdf_compare}}
\end{figure}

\begin{figure}[ptb]
\includegraphics[width=0.46\textwidth,angle=0]{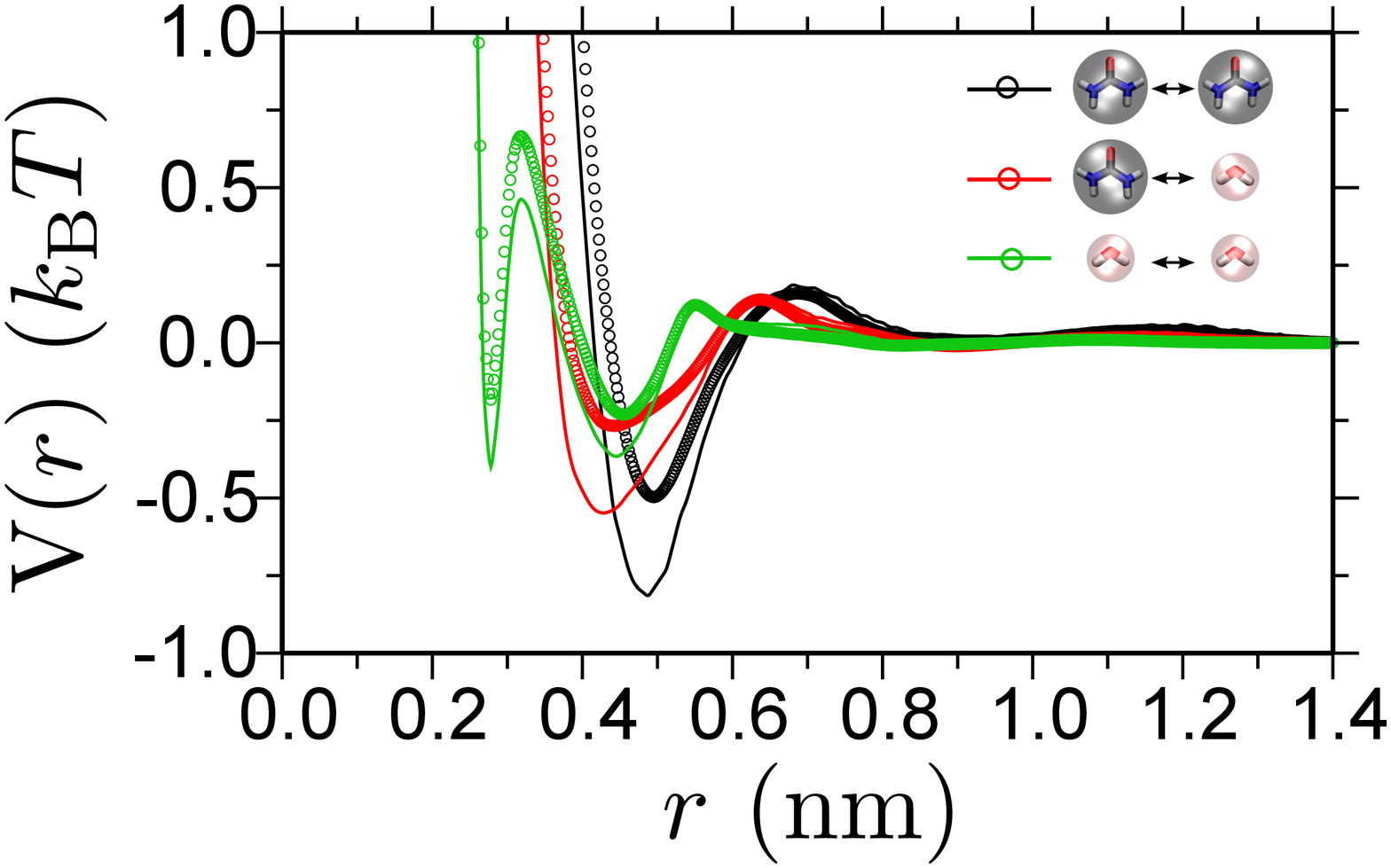}
\caption{Pair-wise coarse-grained (CG) potentials ${{\rm V}(r)}$ between different solvent pairs for a 6 M aqueous urea 
derived using two different CG methods. The solid lines present CG potentials derived from IBI and the symbols are for 
the $\mathcal C-$IBI protocol. Data is shown after 25 iterations within both protocols. 
\label{fig:pot}}
\end{figure}

Before describing our $\mathcal C-$IBI method, we first briefly comment on the conventional IBI method. The procedure starts from an
initial guess for the potential of the CG model using ${\rm g}_{ij}({r})$ obtained from the
all atom simulation,
\begin{equation}
{\rm V}_0(r) = -k_{\rm B} T ~{\ln \left[{\rm g}_{ij}(r)\right]}.
\label{eq:ig}
\end{equation}
Then the potential is updated over several iterations using the protocol,
\begin{equation}
{\rm V}_n^{\rm IBI}(r) = {\rm V}_{n-1}^{\rm IBI}(r) + k_{\rm B} T ~ {\ln \left[\frac {{\rm g}_{ij}^{n-1}(r)} {{\rm g}_{ij}^{\rm target}(r)}\right]}.
\label{eq:ibi}
\end{equation}
During every iteration, a 1 ns long MD run of the CG system is performed using the potential obtained at the end of the preceeding iteration.
In Fig.~\ref{fig:rdf_compare}(a) we present a comparison between fitted ${\rm g(r)}$ after 25 IBI iterations (symbols) and the 
reference all-atom data (solid lines).
At a first look it appears to be in reasonably good agreement. Moreover, the first shell coordination 
${\mathcal C}_{ij}^{f} = 4\pi \int_0^{r_0} {\rm g}_{ij}(r') r'^2 dr'$ shows a deviation of roughly $\sim 2 - 4\%$. Note that
${\mathcal C}_{ij}^{f}$ requires integration over the first peak of ${\rm g}(r)$, thus we have chosen 
$r_0 = 0.32$ nm for water-water, 0.48 nm for urea-water and 0.58 nm for urea-urea distributions, respectively. 
For example, a small error within the first few solvation shells (as observed in ${\rm g}(r)$) cumulatively 
adds up to a large error at the tail and thus severely disturbs particle fluctuations. 
In this context, this small error is not recognized within the IBI protocol, where the 
corrections are weighted with a factor of $1/r^2$ $-$ when looking into the coordination numbers. 
This leads to a position dependent error, which is most severe for large $r$ values, and also added 
cummulative error from the earlier $r$ values. Therefore, there is a need of a protocol, especially for 
binary mixtures, that gives precise solvation properties. A theory that can serve as an 
excellent guide to achieve this purpose is the fluctuation theorem of Kirkwood and 
Buff (KB) \cite{kb51jcp}. KB theory connects the pair-wise coordination with particle fluctuations and, thus, 
with the solution thermodynamics. KB theory makes use of the ``so called" Kirkwood-Buff integrals (KBI) $G_{ij}$ defined as,
\begin{equation}
\begin{split}
{\rm G}_{ij} & = V \left[\frac{\left<N_{i} N_{j}\right> -
\left<N_{i}\right> \left<N_{j}\right>}{\left<N_{i}\right> \left<N_{j}\right>}
 - \frac {\delta_{ij}}{\left<N_{j}\right>}\right] \\
&  = 4\pi \int_0^\infty \left[ {\rm g}_{ij}^{\mu {\rm V T}}(r) - 1\right] r^2 dr \\
\end{split}
\label{eq:kbi1}
\end{equation}
where averages in the grand canonical ensemble ($\mu {\rm VT}$) are denoted by brackets $\left< \cdot \right>$, $V$ is the volume, $N_{i}$
the number of particles of species $i$, $\delta_{ij}$ is the Kronecker delta, and ${\rm g}_{ij}^{\mu {\rm V T}}(r)$
is the pair distribution function in the $\mu {\rm VT}$ ensemble. 
For finite systems, however, a reasonable approximation leads to ${\rm g}_{ij}^{\mu {\rm V T}}(r) \approx {\rm g}_{ij}^{{\rm N V T}}(r)$ 
with ${\rm g}_{ij}^{{\rm N V T}}(r)$ being the pair distribution function in the canonical (NVT) ensemble. 
For big system sizes this is nearly almost (always) a safe approximation and thus leading to
\begin{equation}
\begin{split}
{\rm G}_{ij} (r) &  = 4\pi \int_0^r \left[ {\rm g}_{ij}^{{\rm N V T}}(r') - 1\right] r'^2 dr' \\
&  = {\mathcal C}_{ij}(r) - \frac{4}{3}\pi r^3.
\end{split}
\label{eq:kbi2}
\end{equation}
Here the second term in the last line is a volume correction to ${\mathcal C}_{ij}(r)$. Therefore, the quantity $G_{ij}(r)$
is also refered to as the excess-coordination, which could be connected to solvation properties of 
multi-component mixtures \cite{mukherji13mac,smith03jpcb,naimbook}. Therefore, we not only need the precise estimate of 
${\rm g}(r)$, but also correct $G_{ij}$. This presents a need for an improved protocol that can correctly reproduce pair-wise coordination and 
the solvation properties. Thus we propose coordination iterative Boltzmann inversion ($\mathcal C-$IBI). 
Here also, the initial guess is the same as in Eq.~\ref{eq:ig}. However, the iterative protocol is modified to 
target ${\mathcal C}_{ij}(r)$ given by,
\begin{equation}
{\rm V}_n^{{\mathcal C-}{\rm IBI}}(r) = {\rm V}_{n-1}^{{\mathcal C-}{\rm IBI}}(r) + 
k_{\rm B} T ~ {\ln \left[\frac {{{\mathcal C}_{ij}^{n-1}(r)}} {{{\mathcal C}_{ij}^{\rm target}(r)}}\right]}.
\label{eq:c_ibi}
\end{equation} 
A cut-off distance for ${\mathcal C}_{ij}(r)$ is chosen to be $1.5$ nm, which is typically of the order of the 
correlation length of water-based molecular fluids. 
The advantage of using Eq.~\ref{eq:c_ibi}, unlike the IBI protocol, is that it presents equal weightage at every $r$ value 
and, therefore, corrects ${\mathcal C}_{ij}(r)$ at every $r$ points precisely. Furthermore, because 
${\mathcal C}(r)$ is exactly reproduced using Eq.~\ref{eq:c_ibi}, it also exactly reproduces ${\rm g}(r)$. 
In Fig.~\ref{fig:rdf_compare}(b) we present ${\rm g}(r)$ obtained using the $\mathcal C-$IBI protocol. 
While there is hardly any visible distinction between ${\rm g}(r)$ obtained from $\mathcal C-$IBI and the 
reference all-atom simulations, we find a much improved first shell coordination that shows $\sim0.5\%$ deviation and 
also an improved tail convergence. A comparison of CG potentials derived from both methods, IBI and $\mathcal C-$IBI, is shown in Fig.~\ref{fig:pot}. 
It can be appreciated that the potentials derived from the two methods are distinctly different even when they show very
similar ${\rm g}(r)$ (see Fig.~\ref{fig:rdf_compare}), suggesting that a mere 25 IBI iterations may not be sufficient to 
get the correct coordination and, hence, the solvation properties. 

It should be noted that the IBI protocol is the simplest form of CG method that works exceedingly well for 
several systems \cite{originibi1,originibi2,ibi,wang09epje}. The inital guess of $V(r)$ in IBI is deduced from the Boltzmann
distribution and the subsequent corrections in Eq.~\ref{eq:ibi} are based on the difference in the distribution function 
while ignoring the higher order correlations. Furthermore, IBI can also be considered as IMC without cross-correlation. 
In this context, IMC \cite{imc} can be derived from a thermodynamic argument. 
In our $\mathcal C-$IBI method, we choose the same initial guess as the IBI (see Eq.~\ref{eq:c_ibi}) and 
subsequent corrections are based on the difference in ${\mathcal C} (r)$. 
Because of the nature of $\mathcal C-$IBI protocol, which aim to reproduce ${\mathcal C} (r)$, this also 
tunes any irregularities that may cumulatively add up to an error at large $r$ values. 
Therefore, reproducing ${\mathcal C} (r)$ automatically guarantees the reproduction of underlying ${\rm g} (r)$.
However, just targeting ${\rm g} (r)$ in an iterative procedure may not give a precise estimate of ${\mathcal C} (r)$ and thus may lead to 
unrealistic fluctuation, especially for the multi-component fluids.

\begin{figure}[ptb]
\includegraphics[width=0.46\textwidth,angle=0]{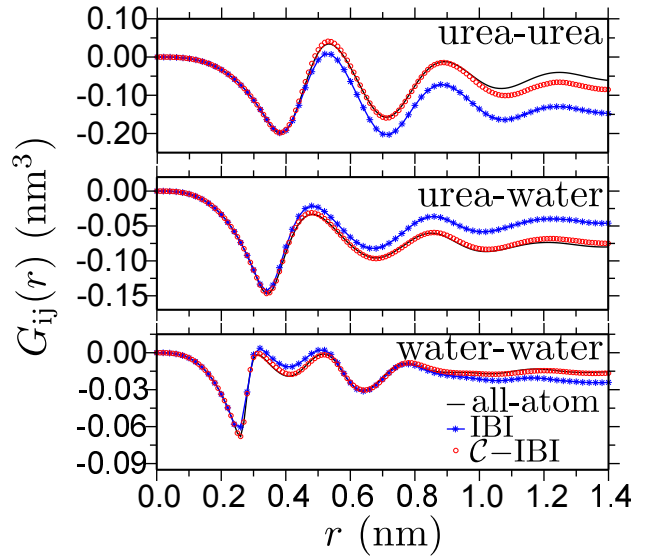}
\caption{Kirkwood-Buff integrals ${G_{ij}(r)}$ between different solvent pairs for a 6 M aqueous urea mixture. 
The all-atom data is compared to the IBI and $\mathcal C-$IBI methods. The data is shown for a 25 ns long MD 
trajectory. The CG simulations are performed with the potentials obtained in Fig.~\ref{fig:pot}. 
\label{fig:kbi}}
\end{figure}

In Fig.~\ref{fig:kbi} we present a comparative plot of $G_{ij}(r)$
between different solvent pairs.
It can be seen that $\mathcal C-$IBI shows a reasonably satisfactory convergence to the reference all-atom data, while IBI data
shows significant deviation, especially between urea$-$urea and urea$-$water. Note that the values of $G_{ij}$ are calculated 
by taking the averages of $G_{ij}(r)$ between 1 nm and 1.4 nm. 
\begin{figure}[ptb]
\includegraphics[width=0.46\textwidth,angle=0]{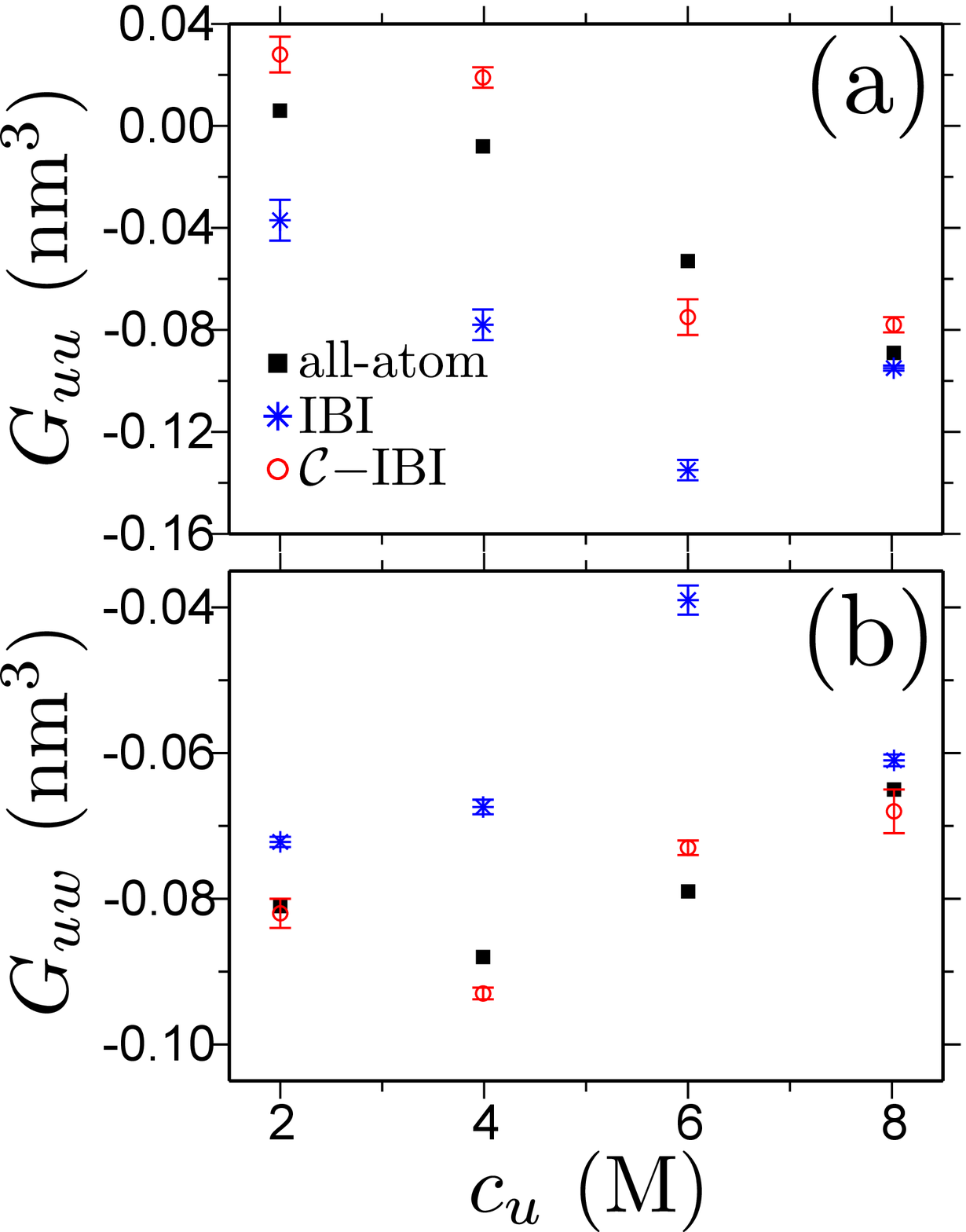}
\caption{Kirkwood-Buff integrals between urea-urea $G_{uu}$ (part a) and urea-water $G_{uw}$ (part b) as a function of molar 
concentration of urea $c_u$. We present a comparative plot of all-atom simulation, IBI and $\mathcal C-$IBI methods. 
The data is shown for a 25 ns long MD trajectory. The error bars are standard deviations obtained from four simulation 
trajectories. Note that we only show $G_{ij}$ between the minor components of urea-urea and urea-water pairs 
that are most effected by the CG protocol. 
\label{fig:kbi_error}}
\end{figure}
In Fig.~\ref{fig:kbi_error} we show $G_{ij}$ between urea-urea $G_{uu}$ and urea-water $G_{uw}$.
It can be appreciated that the data from $\mathcal C-$IBI CG model can closely reproduce $G_{ij}$ obtained from all-atom
simulations. Note that we only show the data for urea-urea and urea-water pairs, where urea is the minor species. For 
water-water KBI, both models give reasonable agreement.
Here it is important to mention that a slight deviation of $G_{ij}$ can result in a large deviation in the particle fluctuation and thus leading to 
wrong solvation thermodynamics. Therefore, in the next section, we will show that our method also gives a correct estimate of 
the solvation free energy.

\begin{table*}[ptb]
\caption{A table showing comparative detail of the first shell excess coordination,
obtained from all-atom (AA) simulations, iterative Boltzmann inversion (IBI) and coordination iterative Boltzmann inversion ($\mathcal C-$IBI).
Results for IBI and $\mathcal C-$IBI are shown for $N_{\rm iterations}$ iterations. For 2 M and 4 M due to the lower urea concentrations
we perform a set of 10 IBI iterations before a set of $\mathcal C-$IBI iterations. This specific choice is adequate to obtain a
reasonably good first estimate of the potential, before starting $\mathcal C-$IBI. 
We also include data for a set of simulations where we use IBI derived CG potential after 125 iterations.}
\label{tab:1st_ex}
\begin{center}
\begin{tabular}{cc|c|c|c|c|c|c|c|c|c|c|c|c|}
\cline{3-14}
                            &               & \multicolumn{12}{c|}{First shell excess coordination = $4\pi\int_0^{r_0} {\rm g}(r') r'^2dr' - 4\pi r_0^3/3$}          \\ \cline{3-14} 
                            &               & \multicolumn{4}{c|}{urea-urea ($r_0 = 0.58$ nm)}        & \multicolumn{4}{c|}{urea-water ($r_0 = 0.48$ nm)}       & \multicolumn{4}{c|}{water-water ($r_0 = 0.32$ nm)}     \\ \hline
\multicolumn{1}{|c|}{$c_u$(M)} & $N_{\rm iterations}$ & AA     & ${\mathcal C}$-IBI   & IBI     & IBI-125 &   AA   & ${\mathcal C}$-IBI   & IBI     & IBI-125 & AA      & ${\mathcal C}$-IBI   & IBI    & IBI-125 \\ \hline
\multicolumn{1}{|c|}{2}     & 10+64         & 0.045  & 0.048              & 0.032   & 0.098   & -0.036 & -0.035  & -0.034  & -0.038  & -0.011  & -0.023  & -0.026 & -0.024  \\ \hline
\multicolumn{1}{|c|}{4}     & 10+30         & 0.011  & 0.023              & -0.009  & -0.010  & -0.033 & -0.030  & -0.030  & -0.030  & -0.005  & -0.005  & -0.004 & -0.005   \\ \hline
\multicolumn{1}{|c|}{6}     & 25            & -0.013 & -0.009             & -0.033  & -0.021  & -0.034 & -0.032  & -0.021  & -0.026  & -0.002  & -0.001  & 0.004  & 0.002  \\ \hline
\multicolumn{1}{|c|}{8}     & 15            & -0.039 & -0.033             & -0.029  & -0.030  & -0.018 & -0.017  & -0.015  & -0.015  &  0.011  &  0.012  & 0.012  & 0.010  \\ \hline
\end{tabular}
\end{center}
\end{table*}

\begin{table*}[ptb]
\caption{Same as table \ref{tab:1st_ex}, but for Kirkwood-Buff integrals (or excess coordination) $G_{ij}$.}
\label{tab:ex_coor}
\begin{center}
\begin{tabular}{cc|c|c|c|c|c|c|c|c|c|c|c|c|}
\cline{3-14}
                            &               & \multicolumn{12}{c|}{$G_{ij}$}                                                                                             \\ \cline{3-14} 
                            &               & \multicolumn{4}{c|}{urea-urea}        & \multicolumn{4}{c|}{urea-water}       & \multicolumn{4}{c|}{water-water}      \\ \hline
\multicolumn{1}{|c|}{$c_u$(M)} & $N_{\rm iterations}$ &    AA   & ${\mathcal C}$-IBI   & IBI     & IBI-125 & AA      & ${\mathcal C}$-IBI   & IBI     & IBI-125 & AA      & ${\mathcal C}$-IBI   & IBI     & IBI-125 \\ \hline
\multicolumn{1}{|c|}{2}     & 10+64         &  0.006  &  0.028  &-0.037    &  0.288  & -0.081  &  -0.082 & -0.072  & -0.105  & -0.023  & -0.023  & -0.024  & -0.020  \\ \hline
\multicolumn{1}{|c|}{4}     & 10+30         & -0.008  &  0.019  &-0.078    &  -0.092 & -0.088  &  -0.093 & -0.067  & -0.067  & -0.015  & -0.015  & -0.021  & -0.020        \\ \hline
\multicolumn{1}{|c|}{6}     & 25            & -0.053  & -0.075  &-0.135    & -0.076  & -0.079  & -0.073  & -0.039  & -0.073  & -0.015  & -0.016  & -0.022  & 0.012  \\ \hline
\multicolumn{1}{|c|}{8}     & 15            & -0.089  & -0.078  &-0.095    & -0.099  & -0.065  & -0.068  & -0.061  & -0.057  & -0.009  & -0.007  & -0.012  &-0.012  \\ \hline
\end{tabular}
\end{center}
\end{table*}

The summary of the first shell excess coordination and the $G_{ij}$ is presented in tables \ref{tab:1st_ex} and \ref{tab:ex_coor} obtained from 
IBI and $\mathcal C-$IBI CG simulations and their comparison to the reference all-atom data. It can be seen that 
for the same number of iterations of both protocols, $\mathcal C-$IBI gives much better estimates of the quantities than the 
standard IBI. It should be noted that the $G_{ij}$ and ${\mathcal C}_{ij}^{f}$ are related to the volume around 
a given molecule. Therefore, smaller molecules also lead to smaller $G_{ij}$ and ${\mathcal C}_{ij}^{f}$ 
values, making them highly sensitive to simulation protocol. Considering this, our $\mathcal C-$IBI method seems to be working 
exceedingly well for the fluid mixtures. Furthermore, $\mathcal C-$IBI also shows faster convergence 
than the IBI protocol. For the 6 M aqueous urea mixture (see Fig.~\ref{fig:kbi}), we get a reasonable convergence within 25 iterations 
of $\mathcal C-$IBI, which otherwise is not possible even after 125 iterations of IBI. In tables \ref{tab:1st_ex} and \ref{tab:ex_coor}
we also include IBI data after 125 iterations. A careful look on the tables also shows that neither $G_{ij}$ nor the 
first shell excess coordination is correctly reproduced within the IBI protocol irrespective of the number of iterations, 
certainly not both quantities at the same time. However, $\mathcal C-$IBI almost, always reproduces both these quantities 
within reasonable accuracy. 

Furthermore, it should also be noted that for the smaller concentrations of urea, 
namely for 2 M and 4 M, we first run a set of 10 iterations of IBI, with 1 ns each step, followed by a certain number 
of $\mathcal C-$IBI iterations. This procedure was performed to obtain a reasonable guess for the initial potential in Eq.~\ref{eq:ig}, 
especially for the urea-urea pairs. Note that the convergence of ${\rm g} (r)$ for large $r$ values are highly sensitive for 
multi-component systems, especially when one of the solvent components present at low concentrations \cite{mukherji13mac,mukherji12jctclet}.
Additionally, we also want to point out that for the smallest urea concentrations, namely 2 M and 4 M, IBI almost never gives any 
reasonable estimate of $G_{ij}$ and ${\mathcal C}_{ij}^{f}$. For example, in tables II and in Fig.~\ref{fig:kbi_error}, it 
can be appreciated that the urea-urea and urea-water KBIs using IBI CG models show large deviations from their all-atom data. 
Thus suggesting that IBI, despite giving after some iterations a reasonable starting potential guess, may still not be a suitable 
scheme to obtain reasonable fluctuations, especially when one of the solvent components are in low concentrations.

We would also like to point out that the $\mathcal C-$IBI CG potentials are obtained without incorporating any adjustable pre-factors 
in the second term of Eq.~\ref{eq:c_ibi}. There are related methods that aim to reproduce KBI of binary \cite{ganguly12jctc} 
and ternary mixtures \cite{ganguly13jctc}. This method makes use of a pressure-like \cite{ibi} KBI-based ramp correction. The
advantage of ramp correction protocol is that it can be used to tune any thermodynamic
property within a simplified protocol, such as the pressure, KBI and/or surface tension.
However, a ramp correction usually requires a careful tuning of the pre-factor. Furthermore,
while the ramp corrections can be used to tune a particular property of interest, it often
sacrifices other properties. For example, when pressure corrections are applied to a system,
it sacrifices fluid compressibility \cite{ibi}. Therefore, the parameter free $\mathcal C-$IBI method is a protocol that,
by construction, reproduces coordination, excess coordination, pair-wise solution structure,
and thus the solvation free energies.
Furthermore, because of the structure based nature of the $\mathcal C-$IBI method, transferability is almost impossible over 
a wide range of concentrations. 
This is because when CG potentials are derived at two concentrations of urea, then these two potentials only give 
precise thermodynamic properties on those two concentration state points. The use of these potentials 
in between concentrations often lead to inconsistent results. Therefore, the structure based CG protocols (such as $\mathcal C-$IBI method) is 
thermodynamically consistent, but presents no concentration transferability. Moreover, when dealing with phase transition 
by changing temperature, one can use the method proposed in Ref. [\onlinecite{mukh12jpcb}] in conjunction with $\mathcal C-$IBI method and 
thus presenting a possibility of obtaining temperature transferable CG model with $\mathcal C-$IBI protocol.
Furthermore, the pressure of the CG model derived using $\mathcal C-$IBI remains around 5000 bars, a typical shortcoming of the almost all CG models. 
In the next section, we will show how a slight change in $G_{ij}$, as reported in the table \ref{tab:ex_coor}, can lead to a large, unphysical, 
deviations in the reference thermodynamic properties. 

\begin{figure}[ptb]
\includegraphics[width=0.46\textwidth,angle=0]{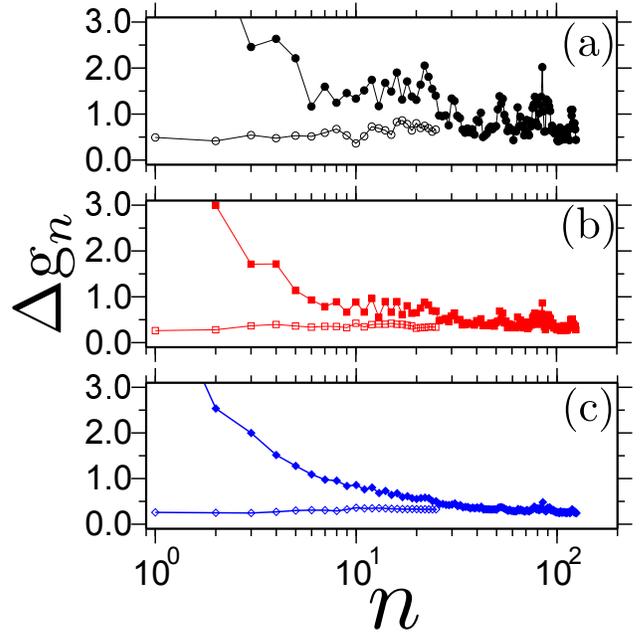}
\caption{Relative error $\Delta {\rm g}_n$ for a 6 M aqueous urea solution obtained over coarse-graining iterations $n$. 
$\Delta {\rm g}_n$ measures the difference between the target radial distribution function ${\rm g}^{\rm target} (r)$ and the 
pair distribution per iterations ${\rm g}_n (r)$, obtained over coarse-graining iterations. Solid 
symbol are obtained from IBI and the open symbols represent $\mathcal C-$IBI. Data is shown for 
urea-urea in part (a), urea-water in part (b) and water-water in part (c). 
\label{fig:converge}}
\end{figure}

Lastly in this section we also want to comment on the convergence of ${\rm g} (r)$ in the $\mathcal C-$IBI scheme and 
the IBI scheme. For this purpose, we calculate the relative error $\Delta {\rm g}_n$ between ${\rm g}^{\rm target} (r)$ and 
${\rm g}_n (r)$ after every iterations $n$, given by
\begin{equation}
\Delta {\rm g}_n = \frac {\sqrt{\int \left[{\rm g}^{\rm target} (r) - {\rm g}_n (r)\right]^2dr}}{\int {\rm g}^{\rm target} (r) dr}.
\label{eq:error}
\end{equation}
In Fig.~\ref{fig:converge} we present $\Delta {\rm g}_n$.
It can be appreciated that the $\mathcal C-$IBI converges much faster than the IBI. 
Furthermore, in both, IBI and $\mathcal C-$IBI corrections, the convergence of one pair always disturbs the convergence 
of others. However, we not only find that $\mathcal C-$IBI converges faster, they also present much less structural fluctuations (see Fig.~\ref{fig:converge}).
More interestingly, we find that a reasonable structure can be obtained from almost very beginning of the $\mathcal C-$IBI protocol, any further 
iterations are performed to get a reasonable convergence of the tail of ${\rm g} (r)$ so that the model can reproduce correct 
fluctuations.

\subsection{Solvation thermodynamics}

\subsubsection{Activity coefficient of aqueous urea}

Solvation of a urea (u) molecule in the mixtures of water (w) and urea can be calculated using the expression \cite{smith03jpcb},
\begin{equation}
\gamma_{uu} = 1 + \left(\frac {\partial \ln \gamma_{u}}{\partial \ln \rho_{u}}\right)_{p,T} 
= \frac{1} {1 + \rho_{u}\left({\mathcal C}_{uu} -{\mathcal C}_{uw}\right)},
\label{eq:solv}
\end{equation}
where $\gamma_{u}$ is the molar cosolvent activity coefficient, $\mu_u = k_{\rm B}T \ln \gamma_{u}$ is the cosolvent chemical potential, 
the urea number density is $\rho_{u}$, ${\mathcal C}_{uu}$ is the urea-urea coordination number and ${\mathcal C}_{uw}$ is the urea-water 
coordination. 
\begin{figure}[ptb]
\includegraphics[width=0.46\textwidth,angle=0]{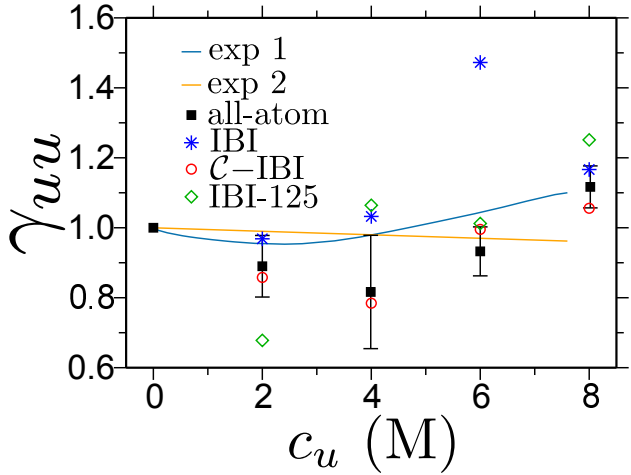}
\caption{$\gamma_{uu}$ as a function of urea molar concentration $c_{u}$ (see Eq.~\ref{eq:solv}). We present comparative data obtained using different coarse-grained method, 
all-atom reference system and experiments. The data set corresponding to experiment 1 is taken from Ref. \cite{stokes67} and experiment 2 is taken from the Ref. \cite{miyawaki97}, 
respectively.
\label{fig:gamma}}
\end{figure}
In Fig.~\ref{fig:gamma} we present $\gamma_{uu}$ as a function of $c_{u}$. The data corresponding to $\mathcal C-$IBI matches 
nicely with the all-atom reference system \cite{mukherji12jctc} and both data sets follow a similar trend as the experimental data set 1 \cite{stokes67}. 
Furthermore, the IBI derived CG models (irrespective of the number of iterations) show a rather random variation in $\gamma_{uu}$. 
Fig.~\ref{fig:gamma} also shows that $\mathcal C-$IBI is a particularly powerful method over the full range of $c_{u}$, while 
the standard IBI CG models only give a slightly better estimate for large $c_{u}$ and for 125 iterations of IBI. 
Note that while the convergence of the tail of ${\rm g}(r)$ is a grand challenge 
within an iterative procedure, $\mathcal C-$IBI appears to be a much better alternative within a reasonable number of iterations.

\subsubsection{Solvation free energy of triglycine in aqueous urea}

So far we have presented results for the aqueous urea mixtures. In this section, we focus on reproducing the solvation properties of a triglycine
in aqueous urea within our $\mathcal C-$IBI protocol. For this purpose, we simulate one triglycine in a box containing water and urea 
with varying $c_{u}$ as described earlier in the method section. For the CG model, we map the full triglycine molecule onto one CG bead. 
Furthermore, as in the cases of 2 M and 4 M aqueous urea mixtures, we first perform an initial set of $25$ IBI iterations, followed by 
30 $\mathcal C-$IBI iterations. This is again motivated by the fact that we want to have a reasonable initial guess for the 
potential (see Eq.~\ref{eq:ig}). Here, however, every iteration consists of a 10 ns MD trajectory. Note that we deliberately chose single triglycine molecules, to test 
the robustness of our method under extreme CG simulation conditions. 

\begin{figure}[ptb]
\includegraphics[width=0.46\textwidth,angle=0]{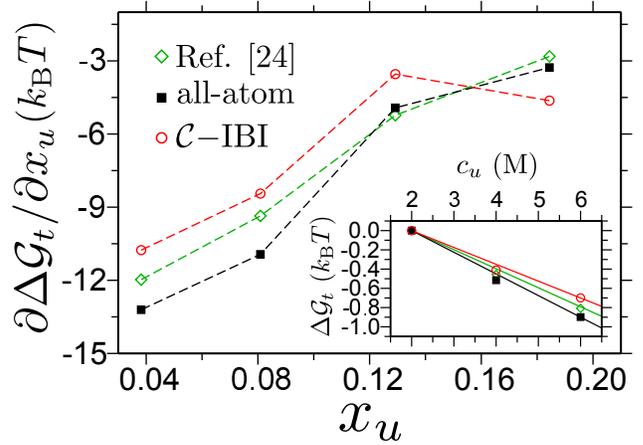}
\caption{Derivative of triglycine solvation free energy ${\partial \Delta \mathcal{G}_{t}}/{\partial x_{u}}$, 
as shown in Eq.~\ref{eq:solv2}, as a function of urea mole fraction $x_u$. Note here we use urea mole fraction, instead of urea molar
concentration $c_u$, in the abscissa to be consistent with the Eq.~\ref{eq:solv2}. We present data for the all-atom simulations and 
from $\mathcal C-$IBI models. For comparison we also include data from Ref.~[\onlinecite{mukherji12jctc}], which was obtained 
using a hybrid multiscale method. In the inset we present the variation of shift in solvation energy of a triglycine $\Delta \mathcal{G}_{t}$
with $c_u$. Note that we only restrict our data in the inset till $c_u =$ 6.0 M concentration of urea, because the experimental data is only available at around 
this $c_u$ value. Straight lines are the linear fits to the data with the slopes listed in table~\ref{tab:mvalue}.
\label{fig:solve}}
\end{figure}

When a triglycine $t$ at infinite dilution ($\rho_t\to 0$) is dissolved in an aqueous urea solution,
the shift in the solvation free energy of triglycine ($\Delta \mathcal{G}_t$) is given by \cite{naimbook},
\begin{equation}
\lim_{\rho_t \rightarrow 0} \left(\frac {\partial \Delta \mathcal{G}_t}{\partial x_u}\right)_{p,T} 
= \frac {k_{\rm B}T \left(\rho_{w} + \rho_{u}\right)^2} {\eta} \left({\mathcal C}_{tw} -{\mathcal C}_{tu}\right),
\label{eq:solv2}
\end{equation}
where $x_u$ is the urea mole fraction, $k_{\rm B}$ is the Boltzmann constant, 
$\eta = \rho_{w} + \rho_{u} + \rho_{w} \rho_{u} \left( {\mathcal C}_{ww} + {\mathcal C}_{uu} - 2{\mathcal C}_{uw} \right)$
is the preferential solvation parameter, and $\rho_i$ is the number density of the $i^{th}$ component of the aqueous solutions. 
In Fig.~\ref{fig:solve} we present ${\partial \Delta \mathcal{G}_t}/{\partial x_u}$ as a function of $x_{u}$.
$\mathcal C-$IBI gives a reasonably good agreement with the all-atom data, while the data corresponding to the IBI 
CG model after 60 iterations did not show any visible convergence of ${\mathcal C}_{tw}$ and ${\mathcal C}_{tu}$ that could be used to 
obtain a reasonable estimate of the solvation energy. Furthermore, we do not only get a reasonable estimate of $\partial \Delta \mathcal{G}_t/ \partial x_u$,
but also for different $G_{ij}$ components in Eq.~\ref{eq:solv2}, i.e., ${\mathcal C}_{tw}$, ${\mathcal C}_{tu}$, and $\eta$.

Integration of Eq.~\ref{eq:solv2} gives the direct measure of the shift in solvation energy $\Delta \mathcal{G}_t$ 
with urea concentration. In the inset of Fig.~\ref{fig:solve} we show the variation of $\Delta \mathcal{G}_t$ with $c_u$. 
The slope of the linear fit to the data in the inset of Fig.~\ref{fig:solve} gives the direct measure of the ``so called" $m-$value,
which is defined as ${\partial \Delta \mathcal{G}_t}/{\partial c_u}$.
Additionally, the $m-$value can be efficiently used to make a reasonable comparison between simulation and experimental observations. 
In table~\ref{tab:mvalue}, we present $m-$values of a triglycine obtained from different methods.
\begin{table}[ptb]
\caption{A comparative table showing $m-$value$= {\partial \Delta \mathcal{G}_t}/{\partial c_u}$ obtained from the linear fits to the 
data in the inset of Fig.~\ref{fig:solve}. The results are shown for all-atom simulations, $\mathcal C-$IBI model, 
previous simulations \cite{mukherji12jctc}, and experimental data \cite{auton05pnas}. Note that while the experimental data are usually 
presented in kJ mol$^{-2}$L, the energy unit in our simulations is $k_{\rm B}T$. Therefore, for better representability we provide 
the $m-$value in both these units.}
\begin{center}
\begin{tabular}{|c|c|c|}
\hline
 & \multicolumn{2}{|c|}{$m-$value}\\\hline
 & $k_{\rm B}T$ mol$^{-1}$L & kJ mol$^{-2}$L  \\\hline
\hline
All-atom simulation                           & -0.225 &  -0.557  \\
$\mathcal C-$IBI simulation                   & -0.175 &  -0.433  \\
Simulation Ref.~[\onlinecite{mukherji12jctc}] & -0.198 &  -0.492  \\
Experiment Ref.~[\onlinecite{auton05pnas}]    & -0.197 &  -0.489  \\
\hline
\end{tabular}  \label{tab:mvalue}
\end{center}
\end{table} 
A reasonably good agreement is observed between $\mathcal C-$IBI, all-atom simulations and experiments \cite{auton05pnas}
suggest that the can be used for any multi-component complex fluids.

\section{Conclusion}
\label{sec:con}

We have presented a parameter free coarse-graining (CG) strategy for complex mixtures. Our method uses cumulative coordination as a target function within 
an iterative protocol. We name our method $\mathcal C-$IBI. $\mathcal C-$IBI method not only gives a correct estimate of the pair-wise coordination, 
but also by construction gives a good estimate of the solvation thermodynamics. More specifically, our CG method correctly reproduces 
both $-$ energy density within the solvation volume and the local concentration fluctuations. Additionally, $\mathcal C-$IBI 
shows much faster convergence with respect to the standard iterative Boltzmann inversion (IBI). We have used 
$\mathcal C-$IBI derived CG potentials to study aqueous urea mixtures and the solvation of a small peptide 
in aqueous urea. The method presents a new, simplified, CG protocol and thus can be 
further used to study more complex (bio-)macromolecular systems, especially in mixed solvent environment.

\section{Acknowledgments}

We thank Christine Peter and Nico van der Vegt for stimulating discussions.
T.E.O. and P.A.N. acknowledges financial support from CNPq and CAPES from 
Brazilian Government and hospitality at the Max-Planck Institut f\"ur Polymerforschung, where this work 
was performed and generous allocation of computational facilities at the supercomputing center 
of CESUP-UFRGS. C.J. thanks LANL for a Director's fellowship. Assigned: LA-UR-15-28326. LANL is operated by
Los Alamos National Security, LLC, for the National Nuclear Security
Administration of the U.S. DOE under Contract DE-AC52-06NA25396.
We thank Robinson Cortes-Huerto, Tanja Kling, Tristan Bereau, and Torsten St\"uhn for critical reading of the manuscript.
Simulation snapshots in this manuscript are rendered using VMD \cite{schulten}.

\begin{appendix}

\section{$\mathcal C-$IBI as an extension in VOTCA}

\begin{figure}[h!]
\includegraphics[width=0.49\textwidth,angle=0]{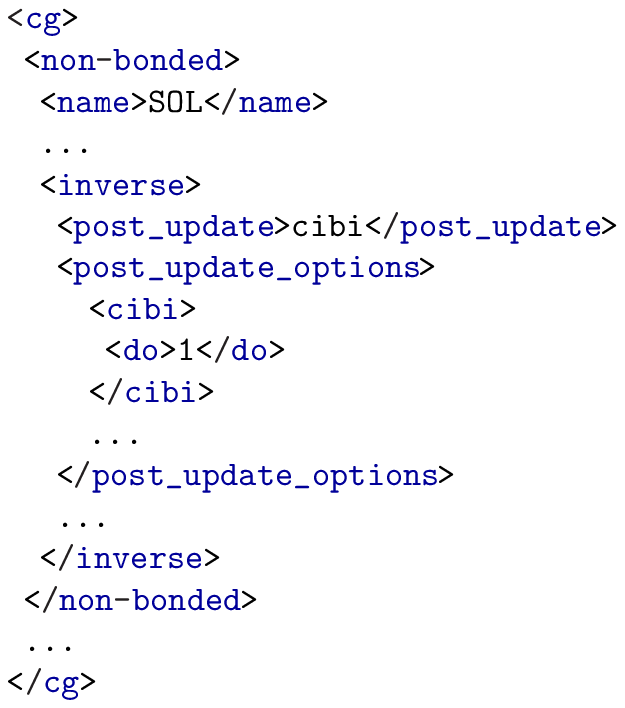}
\caption{A schematic showing part of the script that is required within the settings file for the $\mathcal C-$IBI iterations.
\label{fig:cibi_scr}}
\end{figure}

$\mathcal C-$IBI method is implemented as an extension of the VOTCA package \cite{votca} that requires
certain additional lines, presented in Fig.~\ref{fig:cibi_scr}, to be included within the settings file to perform $\mathcal C-$IBI iterations.

\end{appendix}

\end{document}